\documentclass[english]{article}
\usepackage[T1]{fontenc}
\usepackage[latin9]{inputenc}
\usepackage{amssymb}
\usepackage{graphicx}

\makeatletter
\newcommand{\lyxaddress}[1]{
\par {\raggedright #1
\vspace{1.4em}
\noindent\par}
}

\makeatother

\usepackage{babel}
\begin{document}

\title{\textbf{The commutator algebra of covariant derivative as general
framework for extended gravity. The Rastall theory case and the role
of the torsion}}

\author{\textbf{I. Licata $^{1}$, H. Moradpour $^{2}$ and C. Corda $^{2}$}}
\maketitle

\lyxaddress{\begin{center}
$^{1}$ISEM, Inst. For Scientific Methodology, PA, Italy \& $^{3}$School
of Advanced International Studies on Applied Theoretical and Non Linear
Methodologies in Physics, Bari (Italy) 
\par\end{center}}

\lyxaddress{\begin{center}
\textbf{$^{2}$}Research Institute for Astronomy and Astrophysics
of Maragha (RIAAM), P.O. Box 55134-441, Maragha, Iran
\par\end{center}}
\begin{abstract}
In this short review, we discuss the approach of the commutator algebra
of covariant derivative to analyse the gravitational theories, starting
from the standard Einstein's general theory of relativity and focusing
on the Rastall theory. After that, we discuss the important role of
the torsion in this mathematical framework.

In the Appendix of the paper we analyse the importance of the nascent
gravitational wave astronomy as a tool to discriminate among the general
theory of relativity and alternative theories of gravity.
\end{abstract}
\begin{description}
\item [{Keywords:}] \textbf{Rastall theory of gravity; torsion in gravity;
gravitational waves; commutator algebra; covariant derivative.}
\end{description}

\section{Introduction}

Symmetry and its implications on conservation principles have always
been a virtuous route for research in theoretical physics, starting
from the historical work of Emmy Nother \cite{key-1}. The Yang-Mills
theory \cite{key-2} introduced a new way in order to take into due
account symmetry in physics. In fact, in the Yang-Mills approach,
symmetry is no more a way to look at a physical system. Instead, it
is an heuristic tool which analyses the dynamics of such a physical
system \cite{key-3,key-4}. This idea has an immediate physical meaning
and its origin in classical differential geometry. Let us consider
a highly symmetric and ordered physical structure as a crystal lattice
without stress in a simple Euclidean frame. Every defect, or temperature's
variation, or applied force will modify the structure's equilibrium.
The geometry of the original, stress free, structure will not locally
correspond to the new structure's geometry, and this incompatibility
characterizes the perturbing agent. Coordinates of the new geometry
are not commutative. Hence, each pattern cannot return to its initial
values. This means that the integral operator is not unique and that
the system is not conservative. Thus, the gauging is a compensation's
operation \cite{key-5,key-6}. 

It is possible to analyse these procedures through a general framework
called commutator algebra \cite{key-7,key-8}. This approach is founded
on principles concerning various orders of covariant derivatives'
commutators. The physical meaning is simple, that is, transformations
on the operative manifold (space-time or particle phase space) due
to imposed constraints define the \textquotedbl{}compensation's mechanism\textquotedbl{}
and, in turn, the interaction which one wants to characterize. It
is an approach appropriate to release a formal framework which contextualizes
the various extended gravity theories. In fact, such theories can
be inspected starting from the constraint replacing the non-conservation
of the stress-energy tensor. 

This paper is organized as follows: 
\begin{enumerate}
\item We introduce the key concept of the commutator algebra framework;
\item We consider the constrain on the stress-energy tensor which permits
to re-obtain  the gravitational theory derived by P. Rastall \cite{key-10}.
Rastall was indeed one of the pioneering researchers who questioned
the general relativistic assumption of energy conservation as null
divergence $T_{\mu\nu;\mu}=0$ \cite{key-10}. The Rastall theory
is a particular case of the large framework of ``variations'' of
the general theory of relativity (GTR) which is today largely studied
as potential attempting to solve cosmological problems like dark matter
and dark energy {[}11 - 15{]}. It is interesting to observe that the
Rastall theory cannot be derived through a minimal action principle.
Instead, one derives it under the leading principle of minimal possible
deformation of the standard motion equations \cite{key-10}. 
\item We reanalyse the controversial role of the torsion in gravity in the
framework of the Rastall theory.
\item We insert an appendix discussing the importance of the nascent gravitational
wave (GW) astronomy as a tool to discriminate among the GTR and alternative
theories of gravity.
\end{enumerate}

\section{Compensative Commutator Algebra: an introduction }

That is when the concepts of \emph{local symmetry} and \emph{symmetry
breaking} come into play by fixing the most fecund lines of development
of theoretical physics. In fact, when we impose on the equations of
a physical theory to stay invariant in their form in passing from
a global to a local symmetry, it will be necessary to introduce some
\emph{compensation terms} (in maths jargon ``to make a gauging'')
corresponding to the action of a new \emph{field of forces}. Thus,
the concept of force gets free of the \emph{anthropomorphic flavours}
to become a connection on mathematical spaces. In a bit more formal
terms: a gauge theory is a type of \emph{field theory} in which the
\emph{Lagrangian} (the dynamics of a system) is \emph{invariant} under
a \emph{continuous group} of local transformations. These theories
are called theories of the gauge fields. Global symmetries tell us
something about the observers defined on a substratum, whereas the
local ones describe the interactions and so the dynamics of the entities
living in it.

The natural ``language'' of this kind of theories is differential
geometry. We will try to give an idea of them by using the diagrams
of Category theory, following the exposition in \cite{key-7}, with
a bit of ``elementary lexicon''. Let us consider Figure 1 where
T is an operator defining the passage from X to Y in a suitable, abstract
space or substratum. 
\begin{figure}
\includegraphics[scale=0.7]{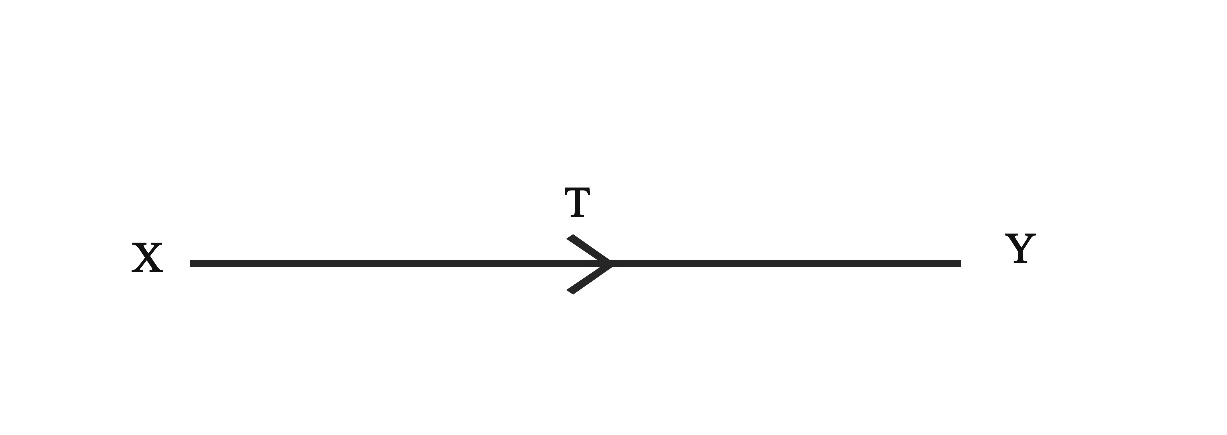}\caption{Connection operators satisfying the commutation relations, adapted
from \cite{key-7}. }
\end{figure}

T can be a global symmetry, or a local one. In the latter case, T
is a gauge transformation. The most important information in the diagram
is that both global and local symmetries must be coherent. The diagram
in Figure 2 contains the conceptual core of gauge theories. In Figure
2 $\Psi$ is a field defined on a substratum and $T(\varphi)$ is
the gauging on the quantity $\varphi$. The directional derivative
$\partial_{k}$ of $\Psi$ is indicated by $\partial_{k}\Psi$, but
the gauging changes it into a more complex expression which implies
a generalization called covariant derivative $D_{k}$. This one is
a connection operator between the spaces defined on the same substratum.
In this way, the term $D_{k}T(\varphi)\Psi$, on the lower right,
closing and guaranteeing the symmetry, indicates the action of a new
field of force linked to $\varphi$ and characterized as a particular
geometrical deformation. 
\begin{figure}
\includegraphics[scale=0.7]{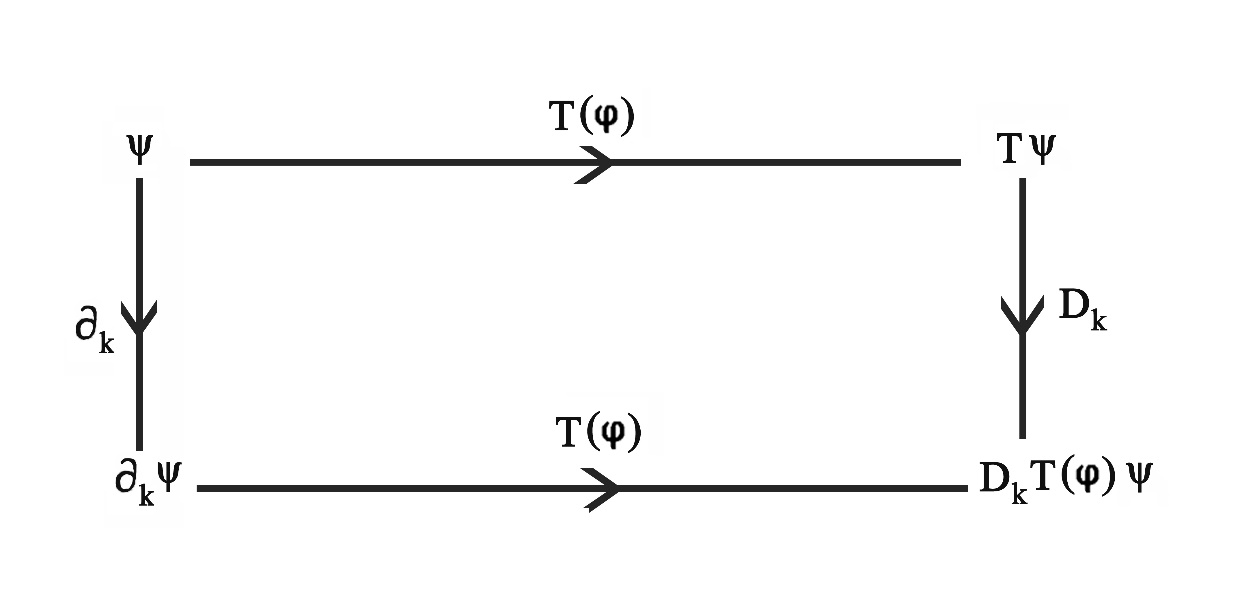}\caption{General form, adapted from \cite{key-7}.}
\end{figure}

Less immediate are the diagrams indicating that the connection operators
must satisfy the \emph{commutation relations} \emph{(or anti-commutation)},
which fix the field potentials and the dynamical equations, see Figure
3. 
\begin{figure}
\includegraphics[scale=0.7]{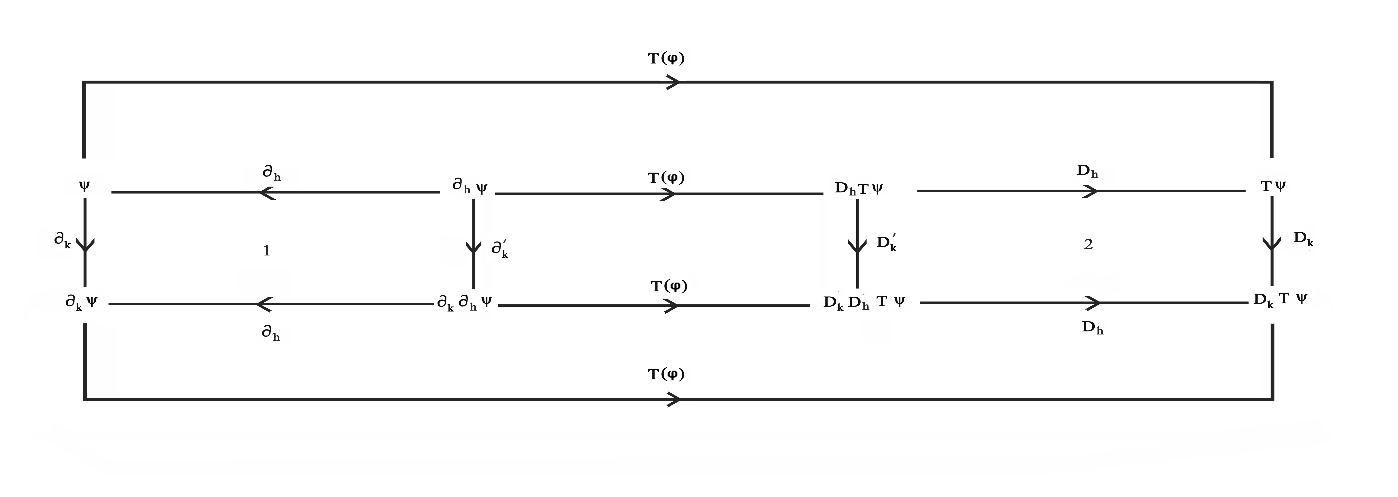}\caption{The core of gauge theories, adapted from \cite{key-7}.}
\end{figure}

Then, if one increases the \emph{Chinese-box}, one gets the most general
form in Figure 4. 
\begin{figure}

\includegraphics[scale=0.7]{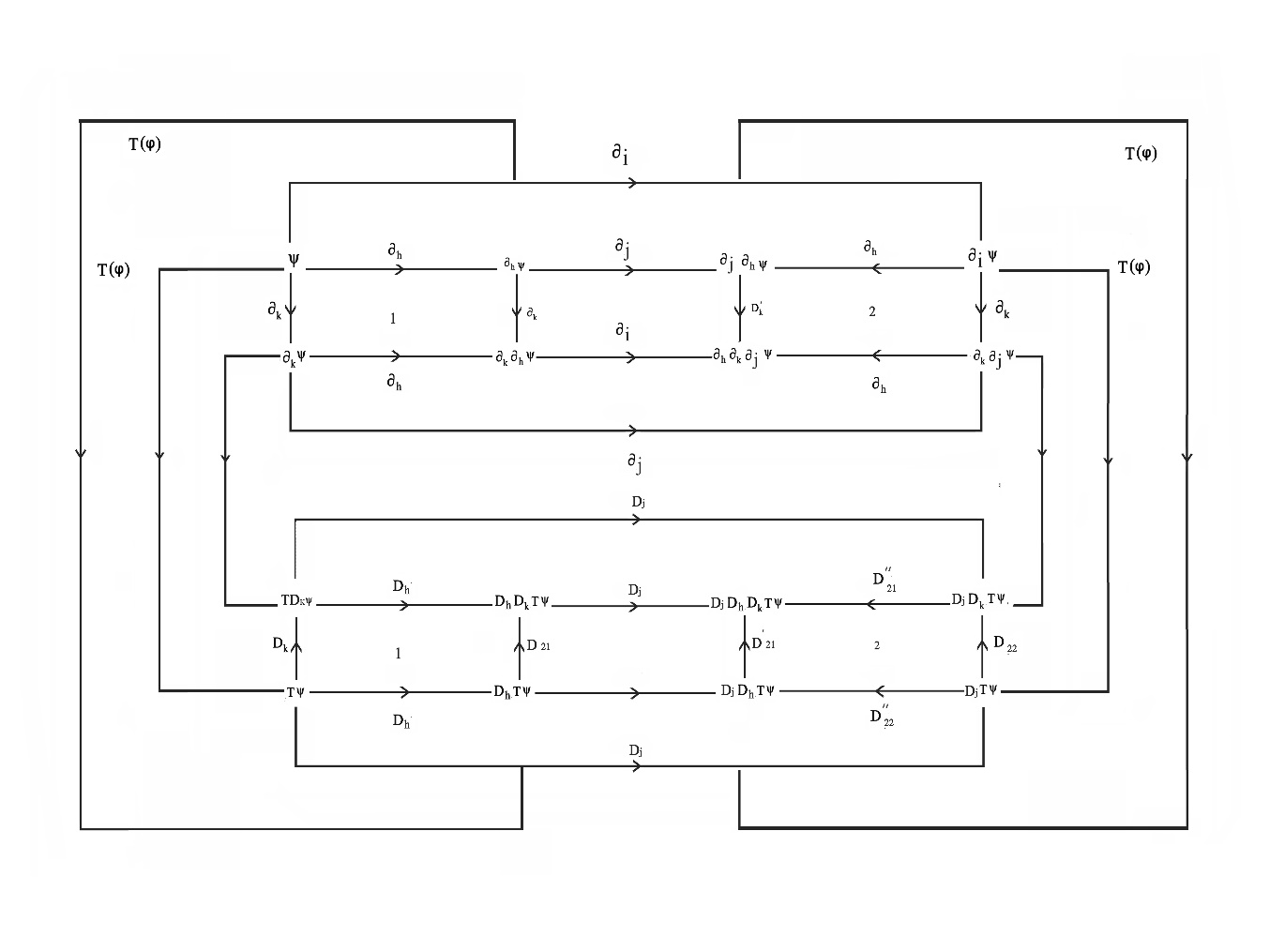}\caption{General form, adapted from\cite{key-7}.}
\end{figure}

Such play of constrains on local symmetries is a mighty ``mathematical
machine'' to build unified theories. By the right gauge conditions
it is actually possible to investigate the relations between different
forces. Obviously, physics is an experimental science and the success
depends on the hypotheses on the substratum and the specific condition
on gauging. We have to do not forget that this scheme, for its very
nature, can give us neither the values of the \emph{fundamental constants}
(such as Planck constant) nor the values of the field source (such
as the electron charge). These are events of the kind $E(x,t)$ and
have to be derived from experience and introduced - as it is used
to say - ``by hand'' in the equations.

Figure 4 suggests the sense of what is meant by unified theory. Each
group can be made up with others, or contained in a bigger group.
The obtained symmetry indicates that interactions had the same intensity
for the value of a certain parameter (for example, temperature in
the standard model), and they differentiate below a critical value
in a \emph{symmetry breaking chain process}. Thus, we can say that
through gauging we look for the tiles of the original symmetry lost
in the history of the Universe. 

Now, let us recall and write down, explicitly, some key formula. Let
us apply the commutator $\nabla_{\mu}$ with the commutator between
$\nabla_{\alpha}$ and $\nabla_{\beta}$ to a vector field $K_{\nu}$.
By introducing the Riemann tensor $R_{\delta\gamma\alpha}^{\beta}$
one writes {[}7 - 9{]}

\begin{equation}
\begin{array}{c}
\left[\nabla_{\mu},\left[\nabla_{\alpha},\nabla_{\beta}\right]\right]K_{\nu}=\\
\\
\left(\nabla_{\mu}\left[\nabla_{\alpha},\nabla_{\beta}\right]K_{\nu}\right)-\left[\nabla_{\alpha},\nabla_{\beta}\right]\left(\nabla_{\mu}K_{\nu}\right)=\\
\\
\left(\nabla_{\mu}R_{\mu\beta\alpha}^{\lambda}\right)K_{\lambda}+R_{\mu\alpha\beta}^{\lambda}\left(\nabla_{\mu}K_{\lambda}\right).
\end{array}\label{eq: Commutator 1}
\end{equation}
It is simple to show that one can obtain the traditional GTR when
the covariant derivative of the vector field $K_{\nu}$ vanishes identically.
Then, from the constraint on the matter source (the ``current'')
$J_{\alpha\beta\gamma}$, which is defined as {[}7 - 9{]}
\begin{equation}
\begin{array}{c}
J_{\alpha\beta\gamma}=D_{\alpha}T_{\beta\gamma}-D_{\beta}T_{\alpha\gamma}-\frac{1}{2}\left(g_{\beta\gamma}D_{\alpha}T-g_{\alpha\gamma}D_{\beta}T\right)\\
\\
for\:which\:D^{\gamma}J_{\alpha\beta\gamma}=0,
\end{array}\label{eq:current}
\end{equation}
where $T_{\alpha\beta}$ is the energy-momentum tensor, $g_{\alpha\beta}$
the metric tensor and $T$ the trace of the energy-momentum tensor,
one gets {[}7 - 9{]}
\begin{equation}
\left[\nabla_{\mu},\left[\nabla_{\alpha},\nabla_{\beta}\right]\right]K_{\nu}=\chi J_{\mu\alpha\beta}K_{\nu}.\label{eq: Commutatore 2}
\end{equation}
Combining eqs. (\ref{eq: Commutator 1}) and (\ref{eq: Commutatore 2}),
one arrives to {[}7 - 9{]}
\begin{equation}
J_{\mu\alpha\beta}K_{\nu}=\left(\nabla_{\mu}R_{\mu\beta\alpha}^{\lambda}\right)K_{\lambda}+R_{\mu\alpha\beta}^{\lambda}\left(\nabla_{\mu}K_{\lambda}\right).\label{eq: arrives}
\end{equation}
 Contracting $\mu$ with $\alpha$ and $\alpha$ with $\nu$ one obtains
the generalized equation of motion {[}7 - 9{]}
\begin{equation}
\nabla_{\mu}\left[R^{\mu\nu}-\chi\left(T^{\mu\nu}-\frac{1}{2}g^{\mu\nu}T\right)\right]K_{\lambda\nu}+R^{\mu\nu}\left(\nabla_{\mu}K_{\lambda}\right)=0,\label{eq: generalized equation of motion}
\end{equation}
Setting $\left(\nabla_{\mu}K_{\lambda}\right)=0$ one gets the Bianchi
identity for a non-null vector field $K_{\nu}$. This is exactly the
case of the GTR {[}7 - 9{]}.

As we previously stressed, eq. (\ref{eq: generalized equation of motion})
does not arises from a minimal action principle on the Lagrangian.
Instead, the added term with respect to the GTR comes from the covariant
derivatives commutator. In particular, the addition's second term
in eq. (\ref{eq: generalized equation of motion}) is the main coupling
term between gravity and the manifold. If this term is null, one recovers
the GTR. Thus, the main difference with the GTR is the origin of the
Riemann tensor from a constraint on the source which results to be
an eigenvalue of the double commutator of covariant derivative. 

With these mathematical tools it is simple introducing the Rastall
theory. This is the object of next Section.

\section{The Rastall theory from commutator algebra}

The Rastall theory is the first example of a theory which considers
a non-divergence-free energy-momentum \cite{key-10}. Another example
is the theory called \emph{the curvature-matter theory of gravity}
{[}16 - 18{]}. In this theory, which is similar to Rastall theory,
the matter and geometry are coupled to each other in a non-minimal
way {[}16 - 18{]}. In that way, the ordinary energy- momentum conservation
law does not work {[}16 - 18{]}. For the sake of completeness, we
stress that these non-minimal theories have been originally introduced
much prior to {[}16 - 18{]} in the pioneer works \cite{key-67,key-68}.

Recent works in the literature renewed the interest in the Rastall
theory, {[}19 - 25{]}. In fact, this theory seems to be in agreement
with observational data on the Universe age and also on the Hubble
parameter \cite{key-26}. In addition, Rastall theory can provide
an alternative description for the matter dominated era with respect
to the GTR \cite{key-27}. It is also supported by observational data
from the helium nucleosynthesis \cite{key-28}. All these evidences
have motivated physicists to study the various cosmic eras in this
framework {[}29 - 33{]}. Indeed, this theory seems to do not suffer
from the entropy and age problems which appear in the standard cosmology
framework \cite{key-34}. Moreover, Rastall theory is also consistent
with the gravitational lensing phenomena \cite{key-35,key-36}. More
studies on this theory can be found in {[}37 - 41{]} and references
therein.

An important point in the more general framework of extended theories
of gravity is that all the potential alternatives to the GTR must
be viable. This means that such alternatives must be metric theories
in order to be in agreement with the Einstein equivalence principle,
which is today supported by a very strong empirical evidence, and
that they must pass the solar system tests \cite{key-42}. Another
key point concerning the viability of extended theories of gravity
is that the recent starting of the gravitational wave (GW) astronomy
with the events GW150914 \cite{key-43} and GW151226 \cite{key-44},
that are the first historical detections of GWs, could be extremely
important in order to discriminate among various modified theories
of gravity. In fact, some differences among such theories can be stressed
in the linearized approximation and, in principle, can be found by
GW experiments, see \cite{key-13} and the Appendix of this review
paper for details. In this context, the analysis of GWs in the Rastall
theory will be the argument of a future work \cite{key-45}. 

In the Rastall's approach the standard energy-momentum tensor is replaced
by an\emph{ effective} \emph{energy-momentum tensor} as \cite{key-10,key-25}
\begin{equation}
T_{\alpha\beta}\longrightarrow S_{\alpha\beta}=T_{\alpha\beta}-\frac{\chi'\lambda'T}{4\chi'\lambda'-1}g_{\alpha\beta},\label{eq: replacement}
\end{equation}
where $\chi'$ and $\lambda'$ are the Rastall gravitational coupling
constant and and the Rastall constant parameter, respectively \cite{key-10,key-25}.
The Rastall constant parameter represents a measure of the tendency
of the geometry (matter fields) to couple with the matter fields (geometry)
leading to the changes into the matter fields (geometry) \cite{key-10,key-25}
. The effective energy-momentum tensor arises from the breakdown of
the ordinary energy-momentum conservation law as \cite{key-10,key-25}
\begin{equation}
T_{;\alpha}^{\alpha\beta}=\lambda'R^{;\beta}.\label{eq: breakdown}
\end{equation}
Thus, matter fields and geometry are coupled to each other in a non-minimal
way in Rastall theory, and one gets compatibility with some observational
data \cite{key-25}. 

The replacement (\ref{eq: replacement}) permits to replace the matter
source with a \emph{Rastall effective} \emph{matter source} as 
\begin{equation}
\begin{array}{c}
J_{\alpha\beta\gamma}\longrightarrow J_{\alpha\beta\gamma}^{Rastall}=D_{\alpha}S_{\beta\gamma}-D_{\beta}S_{\alpha\gamma}-\frac{1}{2}\left(g_{\beta\gamma}D_{\alpha}S-g_{\alpha\gamma}D_{\beta}S\right)\\
\\
for\:which\:D^{\gamma}J_{\alpha\beta\gamma}^{Rastall}=0,
\end{array}\label{eq: replacement 2}
\end{equation}
and $S$ is the trace of the effective energy-momentum tensor. 

Thus, with the same way of thinking of previous Section one gets 
\begin{equation}
\left[\nabla_{\mu},\left[\nabla_{\alpha},\nabla_{\beta}\right]\right]K_{\nu}=\chi'J_{\mu\alpha\beta}^{Rastall}K_{\nu}.\label{eq: Commutatore 3}
\end{equation}
Now, combining eqs. (\ref{eq: Commutator 1}) and (\ref{eq: Commutatore 3}),
one arrives to 
\begin{equation}
J_{\mu\alpha\beta}^{Rastall}K_{\nu}=\left(\nabla_{\mu}R_{\mu\beta\alpha}^{\lambda}\right)K_{\lambda}+R_{\mu\alpha\beta}^{\lambda}\left(\nabla_{\mu}K_{\lambda}\right).\label{eq: arrives 2}
\end{equation}
Then, if one again contracts $\mu$ with $\alpha$ and $\alpha$ with
$\nu$ one obtains a new generalized equation of motion as 
\begin{equation}
\nabla_{\mu}\left[R^{\mu\nu}-\chi'\left(S^{\mu\nu}-\frac{1}{2}g^{\mu\nu}S\right)\right]K_{\lambda\nu}+R^{\mu\nu}\left(\nabla_{\mu}K_{\lambda}\right)=0.\label{eq: generalized equation of motion 2}
\end{equation}
Setting $\left(\nabla_{\mu}K_{\lambda}\right)=0$ one gets the analogous
of the Bianchi identity in the Rastall theory for a non-null vector
field $K_{\nu}$. Also in this case, the addition's second term in
eq. (\ref{eq: generalized equation of motion 2}) is the main coupling
term between gravity and the manifold. Setting this term equal to
zero permits to recover the Rastall theory. Hence, the main difference
between our generalization and the Rastall theory is again the origin
of the Riemann tensor from a constraint on the source which is an
eigenvalue of the double commutator of the covariant derivative. Thus,
also in the current case eq. (\ref{eq: generalized equation of motion 2})
is not generated by a minimal action principle on the Lagrangian,
but the added term with respect to the Rastall theory comes from the
covariant derivatives commutator.

\section{Torsion in the framework of the Rastall theory of gravity}

The Einstein\textendash Cartan\textendash Sciama\textendash Kibble
theory, also called the Einstein\textendash Cartan theory, is the
most famous framework which attempts to take into due account the
presence of the torsion in the gravitational theory {[}46 - 49{]}.
This (classical) theory of gravitation is similar to the GTR but it
has the remarkable difference that it assumes the presence of a torsion
tensor, which works as the vanishing antisymmetric part of the the
affine connection \cite{key-9}. Hence, the torsion is coupled to
the intrinsic spin of matter in analogous way in which the curvature
is coupled to the energy and momentum of matter \cite{key-9}. The
reason is that, if one considers a curved space-time, one sees that
the spin of matter needs torsion to not be null, but, instead, working
as a variable in the variational principle of stationary action Considering
the torsion and metric tensors as being independent variables, the
correct conservation law for the total (orbital plus spin) angular
momentum due to the presence of the gravitational field can be found
\cite{key-9}. From the historical point of view, the theory was originally
developed by E. Cartan from 1922 \cite{key-46} till 1925 \cite{key-47}.
Then, additional contributions came from Sciama \cite{key-48} and
Kibble \cite{key-49}. In his famous search for a unified field theory,
Einstein approached this theory in 1928 through a failed attempt in
which torsion should match the electromagnetic field tensor. Despite
this failure, this attempt led Einstein to the different theory of
\emph{teleparallelism} \cite{key-50}. Today, there are various researchers
who still works on the Einstein\textendash Cartan theory which is
still considered viable, see for example \cite{key-51}.

On the other hand, the potential presence of torsion in gravity is
today considered a controversial issue. For example Hammond claims
that torsion is required for a complete theory of gravity, and that
without it, the equations of gravity violate fundamental laws \cite{key-52}.
Instead, Kleinert thinks that torsion can be moved into the curvature
in an intriguing gauge transformation without changing the physical
content of Einstein\textquoteright s field equations of the GTR \cite{key-53}.
The remarkable consequence is the invisibility of torsion in any gravitational
experiment \cite{key-53}. This is surely an interesting controversy
between two esteemed theoretical physicists.

Following \cite{key-9}, one introduces a new wave equation for gravity
starting from the Riemann tensor which is 
\begin{equation}
R_{\alpha\mu\nu}^{\lambda}\equiv\partial_{\mu}\Gamma_{\nu\alpha}^{\lambda}-\partial_{\nu}\Gamma_{\mu\alpha}^{\lambda}+\Gamma_{\mu\alpha}^{\sigma}\Gamma_{\nu\sigma}^{\lambda}-\Gamma_{\nu\alpha}^{\sigma}\Gamma_{\mu\sigma}^{\lambda}.\label{eq: Riemann tensor}
\end{equation}
By using the ``Maxwellian scheme for gravity'' \cite{key-9} 
\begin{equation}
\begin{array}{c}
R_{\beta\gamma\delta}^{\alpha}+R_{\gamma\delta\beta}^{\alpha}+R_{\delta\beta\gamma}^{\alpha}=0\\
\\
D_{\epsilon}R_{\beta\gamma\delta}^{\alpha}+D_{\gamma}R_{\beta\delta\epsilon}^{\alpha}+D_{\delta}R_{\beta\gamma\epsilon}^{\alpha}=0\\
\\
\left[D_{\gamma},\,\left[D_{\alpha},\,D_{\beta}\right]\right]V_{\delta}=\left(D_{\gamma}R_{\delta\alpha\beta}^{\mu}\right)V_{\mu}+R_{\gamma\alpha\beta}^{\mu}D_{\mu}V_{\mu},
\end{array}\label{eq: Maxwell scheme for gravity 2}
\end{equation}
one uses the third of eqs. (\ref{eq: Maxwell scheme for gravity 2})
as dynamic equation. Now, one connects the commutator with the \emph{effective
Rastall gravity current} as 
\begin{equation}
\chi'J_{\mu\alpha\beta}^{Rastall}=\left(D_{\mu}R_{\nu\alpha\beta}^{\lambda}\right)V_{\lambda}+R_{\mu\alpha\beta}^{\lambda}\left(\nabla_{\lambda}V_{\nu}\right).\label{eq: connect}
\end{equation}
When $\nabla_{\mu}V_{\nu}=0$ one can perform some algebra computation
obtaining the field equations of Rastall theory as 
\begin{equation}
G_{\alpha\beta}=\chi'\left(T_{\alpha\beta}-\frac{\chi'\lambda'T}{4\chi'\lambda'-1}g_{\alpha\beta}\right).\label{eq: Rastall field equations}
\end{equation}
Exactly like the GTR, also the original formulation of the Rastall
theory  was \textquotedbl{}torsion free\textquotedbl{}. Here we add
the torsion to the Rastall framework of gravity. Let us define the
torsion tensor as \cite{key-9}
\begin{equation}
T_{\mu\nu}^{\lambda}\equiv\Gamma_{\mu\nu}^{\lambda}-\Gamma_{\nu\mu}^{\lambda}\label{eq: Torsion}
\end{equation}
One can show that it is possible to show the previous dynamical equation
through a wave equation with particular source being the variables
symmetric and anti-symmetric Christoffel symbols. In that way, torsion
can be inserted in the geometric picture. A wave equation can be introduced
through the calculation of the commutator and of the double commutator.
Thus, we obtain \cite{key-9}
\begin{equation}
D_{\mu}V_{\alpha}\equiv\frac{\partial V_{\alpha}}{\partial x_{\mu}}-\Gamma_{\mu\alpha}^{\lambda}V_{\lambda}.\label{eq: new type of wave}
\end{equation}
Some algebra enables the computation of the first commutator as \cite{key-9}
\begin{equation}
\begin{array}{c}
F_{\mu\nu,\alpha}=\left[D_{\mu},\,D_{\nu}\right]V_{\alpha}=D_{\mu}D_{\nu}V_{\alpha}-D_{\nu}D_{\mu}V_{\alpha}=\\
\\
-\left(R_{\alpha\mu\nu}^{\lambda}V_{\lambda}+T_{\mu\nu}^{\lambda}D_{\lambda}V_{\alpha}\right).
\end{array}\label{eq: first commutator}
\end{equation}
Hence, one obtains \cite{key-9}
\begin{equation}
-F_{\mu\nu,\alpha}=\left[-D_{\mu},\,D_{\nu}\right]V_{\alpha}=G_{\mu\nu,\alpha}+\Omega_{\mu\nu,\alpha}\label{eq: first commutator 2}
\end{equation}
with 
\begin{equation}
\begin{array}{c}
G_{\mu\nu,\alpha}\equiv\left(\frac{\partial\Gamma_{\nu,\alpha}^{\lambda}}{\partial x_{\mu}}-\frac{\partial\Gamma_{\mu,\alpha}^{\lambda}}{\partial x_{\nu}}\right)V_{\lambda}\\
\\
\Omega_{\mu\nu,\alpha}\equiv\left(\Gamma_{\mu,\xi}^{\lambda}\Gamma_{\nu,\alpha}^{\xi}-\Gamma_{\mu,\alpha}^{\xi}\Gamma_{\nu,\xi}^{\lambda}\right)V_{\lambda}+\left(\Gamma_{\mu,\nu}^{\lambda}-\Gamma_{\nu,\mu}^{\lambda}\right)D_{\lambda}V_{\alpha}.
\end{array}\label{eq: imposizioni}
\end{equation}
We use the rescaling \cite{key-9}
\begin{equation}
\Gamma_{\mu,\nu}^{\lambda}\rightarrow\Gamma_{\mu,\nu}^{\lambda}+\frac{\partial\chi}{\partial x_{\mu}}.\label{eq: rescaling}
\end{equation}
$G_{\mu\nu,\alpha}$ is invariant under the rescaling (\ref{eq: rescaling})
\cite{key-9}. Then, by imposing the Lorenz-like gauge condition \cite{key-54}
as \cite{key-9}
\begin{equation}
\frac{\partial\Gamma_{\mu,\nu}^{\lambda}}{\partial x_{\mu}}=0,\label{eq: Lorenz condition}
\end{equation}
a bit of algebra permits to obtain \cite{key-9}
\begin{equation}
\frac{\partial G_{\mu\nu,\alpha}}{\partial x_{\beta}}+\frac{\partial G_{\beta\mu,\alpha}}{\partial x_{\nu}}+\frac{\partial G_{\nu\beta,\alpha}}{\partial x_{\mu}}=0.\label{eq: divergenza nulla}
\end{equation}
One writes down the Lagrangian gravitational density as \cite{key-9}
\begin{equation}
\begin{array}{c}
L\equiv F_{\mu\nu,\alpha}F^{\mu\nu,\alpha}=G_{\mu\nu,\alpha}G^{\mu\nu,\alpha}+\\
\\
\Omega_{\mu\nu,\alpha}\Omega^{\mu\nu,\alpha}+2G_{\mu\nu,\alpha}\Omega^{\mu\nu,\alpha}.
\end{array}\label{eq: Lagrangian gravitational density}
\end{equation}
 $G_{\mu\nu,\alpha}G^{\mu\nu,\alpha}$ and $\Omega_{\mu\nu,\alpha}\Omega^{\mu\nu,\alpha}+2G_{\mu\nu,\alpha}\Omega^{\mu\nu,\alpha}$
represent the Lagrangian density for the free gravitational field
and the reaction field of the vacuum respectively \cite{key-9}. The
gravitation field is connected with the field of the vacuum by the
interaction term \cite{key-9}. The dynamic equations for the Rastall
non conservative gravity can be obtained through some algebra as 
\begin{equation}
\begin{array}{c}
\left[D_{\beta},\,\left[D_{\mu},\,D_{\nu}\right]\right]V^{\alpha}=J_{\mu\nu,\alpha\beta}^{Rastall}\\
\\
D_{\beta}G_{\mu\nu,\alpha}=-J_{\mu\nu,\alpha b}^{Rastall}-D_{\beta}\Omega_{\mu\nu,\alpha}-\left[D_{\mu},\,D_{\nu}\right]D_{\beta}V_{\alpha}
\end{array}\label{eq. dynamic equations}
\end{equation}
with \cite{key-9}
\begin{equation}
D_{\beta}G_{\mu\nu,\alpha}\equiv\frac{\partial G_{\mu\nu,\alpha}}{\partial x_{\beta}}-G_{j\nu,\alpha}\Gamma_{\mu\beta}^{\xi}-G_{\mu j,\alpha}\Gamma_{\nu\beta}^{\xi}-G_{\mu\nu,\xi}\Gamma_{\alpha\beta}^{\xi}.\label{eq: DG}
\end{equation}
Thus, 
\begin{equation}
\frac{\partial G_{\mu\nu,\alpha}}{\partial x_{\beta}}=-J_{\mu\nu,\alpha\beta}-D_{\beta}\Omega_{\mu\nu,\alpha}-\left[D_{\mu},\,D_{\nu}\right]D_{\beta}V_{\alpha}=J_{\mu\nu,\alpha\beta}^{Rastall}.\label{eq: Thus}
\end{equation}
From eq. (\ref{eq: Thus}) we get \cite{key-9}
\begin{equation}
\frac{\partial G_{\mu\nu,\alpha}}{\partial x_{\beta}}=\left(\frac{\partial^{2}\Gamma_{\nu,\alpha}^{\lambda}}{\partial x_{\beta}\partial x_{\mu}}-\frac{\partial^{2}\Gamma_{\mu,\alpha}^{\lambda}}{\partial x_{\beta}\partial x_{\nu}}\right)V_{\lambda},\label{eq: implies}
\end{equation}
and putting $\partial x_{\beta=}\partial x_{\mu}$, one obtains \cite{key-9}
\begin{equation}
\frac{\partial G_{\mu\nu,\alpha}}{\partial x_{\mu}}=\left(\frac{\partial^{2}\Gamma_{\nu,\alpha}^{\lambda}}{\partial^{2}x_{\mu}}-\frac{\partial^{2}\Gamma_{\mu,\alpha}^{\lambda}}{\partial x_{\nu}\partial x_{\mu}}\right)V_{\lambda}.\label{eq: gets}
\end{equation}
From the Lorentz-like gauge we obtain \cite{key-9}
\begin{equation}
\frac{\partial^{2}\Gamma_{\mu,\alpha}^{\lambda}}{\partial x_{\nu}\partial x_{\mu}}=0.\label{eq: condizione}
\end{equation}
Therefore, 
\begin{equation}
\frac{\partial G_{\mu\nu,\alpha}}{\partial x_{\mu}}=\frac{\partial^{2}\Gamma_{\nu,\alpha}^{\lambda}}{\partial^{2}x_{\mu}}V_{\lambda}=\left(\frac{\partial^{2}\Gamma_{\nu,\alpha}^{\lambda}}{\partial^{2}x}-\frac{\partial^{2}\Gamma_{\nu,\alpha}^{\lambda}}{c^{2}\partial^{2}t}\right)V_{\lambda}=J_{\nu\alpha}^{Rastall}.\label{eq: then}
\end{equation}
Setting equal to zero the \emph{Rastall effective gravity currents}
eq. (\ref{eq: then}) becomes
\begin{equation}
\frac{\partial G_{\mu\nu,\alpha}}{\partial x_{\mu}}=\frac{\partial^{2}\Gamma_{\nu,\alpha}^{\lambda}}{\partial^{2}x_{\mu}}V_{\lambda}=\left(\frac{\partial^{2}\Gamma_{\nu,\alpha}^{\lambda}}{\partial^{2}x}-\frac{\partial^{2}\Gamma_{\nu,\alpha}^{\lambda}}{c^{2}\partial^{2}t}\right)V_{\lambda}=0.\label{eq: reduces}
\end{equation}
Thus, the derivative of the Christoffel connection $\Gamma_{\nu,\alpha}^{\lambda}$
has wave behavior and this is analogous to the case of the GTR in
\cite{key-9}.

A bit of algebra permits to write the Rastall effective gravitational
currents as 
\begin{equation}
J_{\mu\nu,\alpha\beta}^{Rastall}=\frac{1}{2}R_{\mu\nu,\alpha\beta}\;\Rightarrow J_{\mu\nu,\alpha\beta}^{Rastall}=R_{\mu\nu,\alpha\beta}-J_{\mu\nu,\alpha\beta}^{Rastall}.\label{eq: =00201Cgravitational currents}
\end{equation}
The second derivatives of the Riemann tensor in eq. (\ref{eq: =00201Cgravitational currents})
represent a reaction of a \emph{Rastall effective virtual matter}
or \emph{Rastall effective medium (vacuum)} while the second derivatives
of $J^{Rastall}$ are the \emph{Rastall effective} \emph{currents}
for the matter and are represented by the \emph{Rastall effective
energetic tensor}. The non-linear reaction of the self-coherent system
generates a \emph{Rastall effective} \emph{current. }This effetive
current is due to the complexity of the Rastall gravitational field
and to the non-linear properties of the gravitational waves as 
\begin{equation}
\left(\frac{\partial^{2}\Gamma_{\nu,\alpha}^{\lambda}}{\partial^{2}x}-\frac{\partial^{2}\Gamma_{\nu,\alpha}^{\lambda}}{c^{2}\partial^{2}t}\right)V_{\lambda}=R_{\nu\alpha}-J_{\nu\alpha}^{Rastall}.\label{eq: justifies}
\end{equation}

\section{Conclusion remarks}

The approach of the commutator algebra of covariant derivative has
been discussed in order to analyse the gravitational theories. We
started from the standard GTR and focused on the Rastall theory. After
that, the important role of the torsion in has been analysed in this
mathematical framework.

We recall that, in a physical framework, an efficient mathematical
procedural must correspond to a complete understanding of its meaning.
The provisional ending of our analysis on the commutator algebra of
covariant derivative as general framework suggests that the reasonable
efficacy of symmetry (here we cite, almost verbatim, a famous paper
of Wigner \cite{key-62}) - starting from global symmetry and arriving
to local gauging - is very deeply seated in the foundational choiche
of Science to argue trought equivalence classes rather than through
single events. In that way, the scientific aptitude progressively
becomes a generator of strategies. This is exactly the case of the
gauge theories, where the imposition of symmetry reveals something
about the intercating physical entities. In other words, symmetry
is the most genuine example of what we call ``physical law'' \cite{key-63}.

On one hand, the gauge can be a winning strategy which can give us
new ideas and results, as it is the case of perturbative theories.
On the other han, it can be also a generator of failures. We indeed
recall that the first version of the Yang-Mills theory \cite{key-2}
was physically unreliable. Einstein's GTR is, perhaps, the most beautiful
and fair example of a gauge theory. The ``simplicity'' of the GTR,
which is fittingly celebrated, depends on the ecomomy of its basic
hypotheses. Based on such hypotheses, it is indeed possible triggering
a plurarity of variations which recently obtained a great attention
in the framework of the extended theories of gravity also because
such theories can, in principle, solve some important problem of the
standard cosmology, like dark matter and dark energy {[}11 - 15{]}.
Maybe the nascent GW astronomy \cite{key-43} could cast light on
this issue, see \cite{key-13} and the Appendix of this paper. 

Among the various extended theories of gravity, based on its particular
behaviors, the Rastall theory \cite{key-10} deserves special attention.
Despite Rastall acknowledged that the 1974 discovery of the PSR B1913+16
binary pulsar by Hulse and Taylor \cite{key-65} gave strong, concrete
support to the GTR because the rate of change of its period is correctly
predicted by GW emission \cite{key-66}, he also stressed that this
is a single result and he insisted that \textquotedbl{}we should still
be asking if there are many theories that have the correct post-Newtonian
limit and the right gravitational radiation'' \cite{key-66}. In
next Appendix the importance of the nascent gravitational wave astronomy
will be analysed as a tool to discriminate among the GTR and alternative
theories of gravity. We again recall that we will discuss GWs in the
Rastall theory in a future work \cite{key-45}. We also stress various
important physical situations where, in principle, the Rastall theory
could be important. The Rastall theory is the first example of a theory
which considers a nondivergence-free energy-momentum {[}10{]}. Recently,
it has been shown that observations admit the violation of ordinary
energy-momentum conservation law meaning that the energy-momentum
sources are nondivergence-free tensors in curved spacetimes \cite{key-69}.
Although this result motivates some physicists to consider the cosmological
consequences of this energy conservation violation in f (R,T) gravity
\cite{key-70,key-71}, the idea that the energy-momentum tensor is
not conserved in curved spacetime is coming back to Rastall {[}10{]}.
Recent works in the literature renewed the interest in the Rastall
theory {[}19 - 25{]}. It has been shown that the horizon entropy of
both static and dynamics spacetimes in Rastall theory differ from
that of the GTR {[}19 - 21{]}. It also seems that the effects of Rastall
correction term to the GTR on the structure of Neutron stars, predicted
by the GTR, are not very impressive {[}22{]}. Static solutions in
the presence of a scalar field are studied in {[}23{]}. More solutions
including some different types of black holes in the Rastall framework
have also been presented \cite{key-39,key-40,key-72}. This theory
also admits traversable wormholes which can meet energy conditions
{[}41{]}. Moreover, it has been shown that a generalization of this
theory may describe the cosmos history without needing Dark Energy
for the current phase and an inflaton field for the primary inflationary
era of cosmos {[}25{]}. Based on this generalization {[}25{]}, the
tendency and ability of space-time to couple with baryonic sources,
which fill cosmos in a non-minimal way, could be the origin of the
primary inflationary era and the current accelerating phase of the
Universe expansion {[}25{]}. In fact, the Rastall theory seems to
be in agreement with observational data on the Universe age and also
on the Hubble parameter {[}26{]}. In addition, the Rastall theory
can provide an alternative description for the matter dominated era
with respect to the GTR {[}27{]}. It is also supported by observational
data from the helium nucleosynthesis {[}28{]}. All these evidences
have motivated physicists to study the various cosmic eras in this
framework {[}29 - 33{]}. Indeed, this theory seems to do not suffer
from the entropy and age problems which appear in the standard cosmology
framework {[}34{]}. In addition, the Rastall theory is also consistent
with the gravitational lensing phenomena {[}35, 36{]}. More studies
on this theory can be found in {[}37, 38{]} and references therein.

For the sake of completeness, we signal some recent important works
on the role of torsion in gravitation, differential geometry and cosmology
{[}73 - 75{]}.

\section{Acknowledgements }

The authors thank an unknown referee for useful comments. C. Corda
was funded by the Research Institute for Astronomy \& Astrophysics
of Maragha (RIAAM) under research project No. 1/5237-105.

\section*{Appendix: gravitational theories in the framework of the nascent
gravitational wave astronomy}

The first observation of GWs from a binary black hole (BH) merger
(event GW150914) \cite{key-43}, which occurred in the 100th anniversary
of Albert Einstein's prediction of GWs \cite{key-55}, represented
a cornerstone for science and for gravitational physics in particular.
In fact, it has been the definitive proof of the existence of GWs,
the existence of BHs having mass greater than 25 solar masses and
the existence of binary systems of BHs which coalesce in a time less
than the age of the Universe \cite{key-43}. The event GW150914 was
also the starting of the GW astronomy, a new era in astrophysics and
space sciences with the great hope to discover new, intriguing information
on the Universe. An important point is that the nascent GW astronomy
could be useful in order to discriminate, in an ultimate way, among
the GTR and potential alternative theories. Let us consider, for example,
$f(R)$ theories and scalar tensor gravity (STG), which seem to be
the most popular among gravitational physicists. In fact, they could
be, in principle, important for solving some problem of the standard
cosmology like the dark matter and dark energy problems {[}11 - 15{]}.
$f(R)$ theories and STG attempt to extend the framework of the GTR
through a modification of the Lagrangian, with respect to the standard
Einstein-Hilbert gravitational Lagrangian. In such theories, high-order
terms in the curvature invariants (terms like $R^{2}$, $R^{ab}R_{ab}$,
$R^{abcd}R_{abcd}$, $R\Box R$, $R\Box^{k}R$) and/or terms with
scalar fields non-minimally coupled to geometry (terms like $\phi^{2}R$
) are indeed added to the gravitational Lagrangian {[}11 - 15{]}.
In this Appendix we will focus on these two classes of alternative
theories of gravity. We stress that lots of them can be excluded by
requirements of cosmology and solar system tests {[}11 - 15{]}. Thus,
one needs the additional assumption that the variation from the standard
GTR must be weak \cite{key-13}. 

For the goals of this Appendix the key point is that STG and $f(R)$
theories have an additional GW polarization which, in general, is
massive with respect to the two standard massless polarizations of
the GTR; see {[}13, 56 - 60{]}. One recalls that GW detection is performed
in a laboratory environment on Earth {[}57 - 61{]}. Hence, one typically
uses the coordinate system in which space-time is locally flat and
the distance between any two points is given simply by the difference
in their coordinates in the sense of Newtonian physics \cite{key-61}.
This is the so-called gauge of the local observer {[}57 - 61{]}. In
such a gauge the GWs manifest themselves by exerting tidal forces
on the masses (the mirror and the beam-splitter in the case of an
interferometer) {[}57 - 61{]}. Let us put the beam-splitter in the
origin of the coordinate system. Then, the components of the separation
vector are the coordinates of the mirror {[}57 - 61{]}. The effect
of the GW is to drive the mirror to have oscillations {[}57 - 61{]}.
One considers a mirror having the initial (unperturbed) coordinates
$x_{M0}$, $y_{M0}$ and $z_{M0}$ , where there is a GW propagating
in the $z$ direction. In the GTR the GW admits only the standard
$+$ and $\times$ polarizations \cite{key-61}. Labelling the respective
metric perturbations as $h_{+}$ and $h_{\times}$, to the first order
approximation of $h_{+}$ and $h_{\times}$, the motion of the mirror
due to the GW is \cite{key-61}
\begin{equation}
\begin{array}{c}
x_{M}(t)=x_{M0}+\frac{1}{2}[x_{M0}h_{+}(t)-y_{M0}h_{\times}(t)]\\
\\
y_{M}(t)=y_{M0}-\frac{1}{2}[y_{M0}h_{+}(t)+x_{M0}h_{\times}(t)]\\
\\
z_{M}(t)=z_{M0}.
\end{array}\label{eq: traditional GTR}
\end{equation}
STG has a third additional mode that can be massless \cite{key-57,key-58,key-60}.
In this case, labelling the metric perturbation due to the additional
GW polarization as $h_{\Phi}$ , to the first order approximation
of $h_{+}$, $h_{\times}$ and $h_{\Phi}$, the motion of the mirror
due to the GW is \cite{key-57,key-58,key-60}

\begin{equation}
\begin{array}{c}
x_{M}(t)=x_{M0}+\frac{1}{2}[x_{M0}h_{+}(t)-y_{M0}h_{\times}(t)]+\frac{1}{2}x_{M0}h_{\Phi}(t)\\
\\
y_{M}(t)=y_{M0}-\frac{1}{2}[y_{M0}h_{+}(t)+x_{M0}h_{\times}(t)]+\frac{1}{2}y_{M0}h_{\Phi}(t)\\
\\
z_{M}(t)=z_{M0}.
\end{array}\label{eq: massless STG}
\end{equation}
$f(R)$ theories have a third additional mode which is generally massive
\cite{key-56,key-57,key-59,key-60}. The cases of STG and $f(R)$
theories having a third massive additional mode are totally equivalent
\cite{key-56,key-57,key-59,key-60}. This is not surprising because
it is well known that there is a more general conformal equivalence
between $f(R)$ theories and STG {[}11, 56 - 60{]}. Again, we label
the metric perturbation due to the additional GW polarization as $h_{\Phi}$.
To the first order approximation of $h_{+}$, $h_{\times}$ and $h_{\Phi}$,
the motion of the mirror due to the GW in STG and $f(R)$ theories
having a third massive additional mode is \cite{key-57,key-59,key-60}.
\begin{equation}
\begin{array}{c}
x_{M}(t)=x_{M0}+\frac{1}{2}[x_{M0}h_{+}(t)-y_{M0}h_{\times}(t)]+\frac{1}{2}x_{M0}h_{\Phi}(t)\\
\\
y_{M}(t)=y_{M0}-\frac{1}{2}[y_{M0}h_{+}(t)+x_{M0}h_{\times}(t)]+\frac{1}{2}y_{M0}h_{\Phi}(t)\\
\\
z_{M}(t)=z_{M0}+\frac{1}{2}z_{M0}\frac{m^{2}}{\omega^{2}}h_{\Phi}(t),
\end{array}\label{eq: massive polarization}
\end{equation}
where $m$ and $\omega$ are the mass and the frequency of the GW's
third massive mode, which is interpreted in terms of a wave packet
\cite{key-56,key-57,key-59,key-60}. We also recall that the relation
between the mass and the frequency of the wave packet is given by
\cite{key-56,key-57,key-59,key-60}
\begin{equation}
m=\sqrt{(1-v_{G}^{2})}\omega,\label{eq: relazione massa-frequenza}
\end{equation}
where $v_{G}$ is the group-velocity of the wave-packet. Inserting
eq. (\ref{eq: relazione massa-frequenza}) in the third of eqs. (\ref{eq: massive polarization})
one gets 
\begin{equation}
\begin{array}{c}
x_{M}(t)=x_{M0}+\frac{1}{2}[x_{M0}h_{+}(t)-y_{M0}h_{\times}(t)]+\frac{1}{2}x_{M0}h_{\Phi}(t)\\
\\
y_{M}(t)=y_{M0}-\frac{1}{2}[y_{M0}h_{+}(t)+x_{M0}h_{\times}(t)]+\frac{1}{2}y_{M0}h_{\Phi}(t)\\
\\
z_{M}(t)=z_{M0}+\frac{(1-v_{G}^{2})}{2}z_{M0}h_{\Phi}(t).
\end{array}\label{eq: massive polarization 2}
\end{equation}
The presence of the little mass $m$ implies that the speed of the
third massive mode is less than the speed of light; this generates
the longitudinal component and drives the mirror oscillations of the
$z$ direction \cite{key-56,key-57,key-59,key-60}, which is shown
by the third of eqs. (\ref{eq: massive polarization}). 

The key point here is the following. Only a perfect knowledge of the
motion of the interferometer's mirror will permit one to determine
if the GTR is the definitive theory of gravity. In order to ultimately
conclude that the GTR is the definitive theory of gravity, one must
prove that the oscillations of the interferometer's mirror are in
fact governed by eqs. (\ref{eq: traditional GTR}). Otherwise, if
one proves that the oscillations of the interferometer's mirror are
in fact governed by eqs. (\ref{eq: massless STG}) or eqs. (\ref{eq: massive polarization}),
then the GTR must be extended. 

On the other hand, at the present time, the sensitivity of the current
ground based GW interferometers is not sufficiently high to determine
if the oscillations of the interferometer's mirror are governed by
eqs. (\ref{eq: traditional GTR}), or if they are governed by eqs.
(\ref{eq: massless STG}) or eqs. (\ref{eq: massive polarization}).
A network including interferometers with different orientations is
indeed required and we're hoping that future advancements in ground-based
projects and space-based projects will have a sufficiently high sensitivity.
Such advancements would enable gravitational physicists to determine,
with absolute precision, the direction of GW propagation and the motion
of the various involved mirrors. In other words, in the nascent GW
astronomy we hope not only to obtain new, precious astrophysical information,
but we also hope to be able to discriminate between eqs. (\ref{eq: traditional GTR}),
eqs. (\ref{eq: massless STG}), and eqs. (\ref{eq: massive polarization}).
Such advances in GW technology would equip scientists with the means
and results to ultimately confirm the GTR or, alternatively, to ultimately
clarify that the GTR must be extended. We hope to add to this framework
also the analysis of the oscillations of the interferometer's mirror
due to GWs in Rastall theory in a future work \cite{key-45}. 
\end{document}